\documentclass[12pt]{article}
\usepackage{graphicx}
\usepackage{amsmath}
\usepackage{amsfonts}
\usepackage{overpic}

\newcommand\pubnumber{Article 30 in eConf C1304143}
\newcommand\pubdate{\today}

\def\nyu{Center for Cosmology and Particle Physics, Department of Physics\\
New York University, New York, NY 10003}

\def\Title#1{\begin{center} {\Large #1 } \end{center}}
\def\Author#1{\begin{center}{ \sc #1} \end{center}}
\def\Address#1{\begin{center}{ \it #1} \end{center}}

\newcommand\pubblock{\rightline{\begin{tabular}{l} \pubnumber\\
         \pubdate  \end{tabular}}}
\newenvironment{Abstract}{\begin{quotation}  }{\end{quotation}}
\newenvironment{Presented}{\begin{quotation} \begin{center} 
             PRESENTED AT\end{center}\bigskip 
      \begin{center}\begin{large}}{\end{large}\end{center} \end{quotation}}
\def\Acknowledgements{\bigskip  \bigskip \begin{center} \begin{large}
             \bf ACKNOWLEDGEMENTS \end{large}\end{center}}

\def\beq{\begin{equation}}
\def\eeq#1{\label{#1}\end{equation}}
\def\eeqn{\end{equation}}

\def\beqa{\begin{eqnarray}}
\def\eeqa#1{\label{#1}\end{eqnarray}}
\def\eeqan{\end{eqnarray}}

\let\bar=\overbar

\def\ie{{\it i.e.}}
\def\eg{{\it e.g.}}

\def\Dslash{\not{\hbox{\kern-4pt $D$}}}
\def\dslash{\not{\hbox{\kern-2pt $\del$}}}

\def\msb{{\bar{\ssstyle M \kern -1pt S}}}

\newcommand{\scalefit}{{\texttt{ScaleFit}}}
\newcommand{\boxfit}{{\texttt{BoxFit}}}
\newcommand{\swift}{{\it Swift}}
\newcommand{\ram}{{\texttt{RAM}}}
\newcommand{\emcee}{{\texttt{emcee}}}

\newcommand{\tobs}{t_{obs}}
\newcommand{\Ei}{E_{iso}}
\newcommand{\thO}{\theta_0}
\newcommand{\thobs}{\theta_{obs}}
\newcommand{\epse}{\epsilon_e}
\newcommand{\epsB}{\epsilon_B}
\newcommand{\xN}{\xi_N}
\newcommand{\Fp}{F_{peak}}
\newcommand{\cf}{\mathfrak{f}}
\newcommand{\cfp}{\cf_{peak}}
\newcommand{\cfm}{\cf_{m}}
\newcommand{\cfc}{\cf_{c}}

\begin{document}
\begin{titlepage}
\pubblock

\vfill
\Title{Fitting Afterglows With Multi-Dimensional Simulations}
\vfill
\Author{ Geoffrey Ryan, Hendrik van Eerten, Andrew MacFadyen}
\Address{\nyu}
\vfill
\begin{Abstract}
We present preliminary data fit results of synthetic light curves computed from numerical afterglow blast wave simulations.  Our technique uses Markov chain Monte Carlo (MCMC) in a new data analysis tool, \scalefit{}.  Scaling relations in both the hydrodynamics and radiation equations allow light curves to be parameterized by a small set of scale-invariant characteristic quantities. These quantities have been calculated and tabulated from high resolution two-dimensional hydrodynamic simulations. Producing a light curve from the characteristics takes only a millisecond, allowing for the use of MCMC data fitting techniques which can require millions of iterations. \scalefit{} is a portable, lightweight, python package which performs this analysis on afterglow light curves. Using the set of \swift{}-XRT light curves from 2011 \& 2012 with known redshifts, we find \scalefit{} can measure the jet opening angle, observer angle, and spectral index of most afterglows. Globally we find gamma-ray burst afterglows tend to be observed off axis, at a significant fraction of the jet opening angle.
\end{Abstract}
\vfill
\begin{Presented}
Huntsville Gamma Ray Burst Symposium\\
Nashville TN, USA, April 14-18, 2013
\end{Presented}
\vfill
\end{titlepage}
\def\thefootnote{\fnsymbol{footnote}}
\setcounter{footnote}{0}

\section{Introduction}

Gamma-ray burst (GRB) afterglows present a rich opportunity for the study of GRBs themselves and their extragalactic environments.  The complex nature of afterglow emission requires the use of numerical simulations for accurate construction of light curves from basic physical parameters.  However, since state-of-the-art simulations require days to produce a light curve, it is challenging to use the most accurate simulations in a live data analysis situation.  The \boxfit{} package made use of scaling invariance between explosion energies and circumburst medium densities in the hydrodynamic equations to speed the process, tabulating the results of simulations so only radiative transfer need be performed at run time \cite{vanEerten:2011yn}.  \scalefit{} extends this work, using scale invariance in synchrotron spectra to generate light curves directly from a precomputed table \cite{vanEerten:2011bf}.  These light curves have the accuracy of advanced numeric simulations, but can be generated in milliseconds, opening new afterglow data analysis possibilities. 

\scalefit{} is a python package which implements this scaling procedure to perform Markov-chain Monte Carlo (MCMC) data-fitting on GRB afterglow light curves.  It outputs central values and uncertainties in all fit parameters as well as a list of samples approximating the full posterior probability distribution function (PDF) of the fit.  As a first run we have performed fits for all \swift-XRT afterglows in 2011 and 2012 with known redshifts.  We find \scalefit{} can constrain values for $\thO$, $\thobs$, and $p$ for several bursts, while other parameters are relatively unconstrained by the single-band fit.  Our preliminary results indicate most afterglows are observed significantly off-axis with $p \approx 2.1$. 

\section{ScaleFit}

We model an afterglow as synchrotron radiation produced from a collimated relativistic blast wave propagating through the GRB circumburst medium.  The observed radiation has a synchrotron spectrum parameterized as series of power laws with a peak flux $\Fp$ and break frequencies $\nu_m$ and $\nu_c$ \cite{Sari:1998, Granot:2001ge, vanEerten:2011yn, vanEerten:2011bf}.

For now we ignore self-absorption effects as the corresponding frequency $\nu_a$ lies well-below the \swift{} x-ray band currently under consideration.  Each of these spectral parameters ($\Fp$, $\nu_m$, and $\nu_c$) vary with time and depend on qualities of the blastwave, its environment, and its distance/orientation from the observer.  We parameterize this dependence through the redshift $z$, luminosity distance $d_L$, isotropic-equivalent energy $\Ei$, the circumburst medium density $n_0$, jet half-opening angle $\thO$, observer angle $\thobs$, spectral index $p$, electron energy fraction $\epse$, magnetic energy fraction $\epsB$, and fraction of accelerated particles $\xN$.  We assume a global cooling time and homogenous circumburst medium.  The dependence of the synchrotron spectrum on these parameters is given by simple scaling relations \cite{vanEerten:2011bf}.

After scaling, all the dynamic behaviour of the spectral light curve $F_\nu (\tobs)$ is enclosed in characteristic quantities $\cfp$, $\cfm$, and $\cfc$ which only depend on $\thO$ and $\thobs$.  A series of high resolution, two dimensional numerical simulations have been performed to cover this parameter space using the adaptive-mesh-refinement relativistic hydrodynamics code \ram{} \cite{vanEerten:2011yn, Zhang:2005qy}.  The results are collated into lookup tables for each of the characteristic quantities.  These tables are the core input to \scalefit{}, which can then use the scaling relations to produce light curves for arbitrary values of $\Theta \equiv \{z, d_L, \Ei, n_0, \thO, \thobs, p, \epse, \epsB, \xi_N\}$. 

The \scalefit{} parameter set is ten dimensional and may be highly correlated in some parameters (\eg{} between $\Ei$ and $n_0$, or $\epse$, $\epsB$, and $\xN$) \cite{Eichler:2005ug}.  For these reasons we use MCMC as the core data-fitting routine.  For given data $D$, MCMC produces a set of samples which well approximate the posterior PDF $p(\Theta|D)$.  These samples may then be used to find central values, standard uncertainties, or any other required statistic.  

Our data take the form of light curves: $D = \{ (t_i, F_i, \sigma_{Fi}) \}$.  We take the likelihood function $p(D|\Theta)$ to be a product of independent gaussians for each data point in the light curve.  This gives a likelihood of the standard $\chi^2$ form:
\beq{}
	p(D|\Theta) \propto \exp\left(-\frac{1}{2}\chi^2\right) \ , \qquad \chi^2 = \sum_{i} \left( \frac{F_i - F_{model}(t_i; \Theta) }{ \sigma_{Fi} } \right)^2
\eeq{eq:likelihood}
We take the prior $p(\Theta)$ to be flat within appropriate bounds for each parameter.  For parameters which may vary over several orders of magnitude, the fit is performed (and the prior applied) in log-space.  The bounds used in this analysis are summarized in Table \ref{tb:prior}.

\begin{table}[t] \begin{center} \begin{tabular}{|c|c||c|c|}
\hline
Parameter & Bounds & Parameter & Bounds \\
\hline
$z$ & $[0.0,10.0]$ & $\log_{10} (d_L / 10^{28} \text{cm})$ & $[-5.0,5.0]$ \\
$\log_{10} (\Ei / 10^{53} \text{erg})$ & $[-7.0,3.0]$ & $\log_{10} (n_0 / 1 \text{cm}^{-3})$ & $[-5.0,5.0]$ \\
$\thO $& $[0.045,0.5]$ & $\thobs / \thO$ & $[0.0,1.0]$ \\
$p$ & $[2.0, 3.0]$ & $\log_{10} \epse$ & [-5.0, 0.0] \\
$\log_{10} \epsB$ & [-5.0, 0.0]  & $\log_{10} \xN$ & [-5.0, 0.0] \\
\hline
\end{tabular}
\caption{Default priors for all parameters in \scalefit{}.  For parameters which vary over several orders of magnitude the fit is performed on the logarithm of the parameter instead of the parameter itself.}
\label{tb:prior}
\end{center} \end{table}

To perform the MCMC analysis \scalefit{} uses the \emcee{} package \cite{ForemanMackey:2012ig}.  \emcee{} is a free, open source, python-based package for performing MCMC.  In particular \scalefit{} uses the \textsf{EnsembleSampler}, an implementation of the affine-invariant ensemble MCMC algorithm \cite{Goodman:2010}.  The performance of this algorithm is invariant under affine transformations, providing efficient sampling of highly correlated parameters (a common problem when using simple Metropolis-Hastings type samplers).  

\section{Dataset}
We performed fits on a sample of afterglow light curves made publicly available by the \swift{} collaboration \cite{Evans:2008wp}.  To reduce the dimensionality of the fits we chose to only examine afterglows with known redshifts $z$ and used a benchmark $\Lambda$CDM cosmology ($\Omega_m=0.27, H_0 = 71$ km s$^{-1}$ Mpc$^{-1}$) to calculate $d_L$.  To further reduce the dimensionality and remove degeneracy between the parameters we fix $\xN = 1$ for all fits in this analysis \cite{Eichler:2005ug}.

	The \scalefit{} afterglow model includes the effects of shock deceleration and spreading but does not include flares, energy injection, or other effects in the early time evolution of the light curve.  As such, these parts of the \swift{} data must be identified (with some confidence) and cut out so that fits are only attempted in the regime where the model applies.  Assuming one of $\nu_c$ or $\nu_m$ lies below the observation band the shallowest power law slope obtainable by \scalefit{} is $-0.25$, and the power law slope monotonically decreases with time \cite{Granot:2001ge}.  Hence we include only late-time, steepening, sections of the data with a power law slope smaller than $-0.25$.  We use automatic pre-analysis data published by \swift{} to make this determination \cite{Evans:2008wp}.  If fewer than ten data points remain after the cut, the afterglow is not fit or included in the sample.
	
	There are 38 \swift{}-XRT afterglows with redshifts in 2011 and 2012.  Of these, 33 pass the cut on number of data points.  The raw data was the count-rate light curve for each burst made available by \swift{}.  This was translated to a intrinsic flux light curve via the published counts-to-flux conversion factors which take into account host extinction and galactic absorption.  To properly fit the data, the fit was performed for the \scalefit{} specific flux integrated over the \swift{}-XRT observation band: 0.3 - 10.0 keV.

\section{Results}

\begin{figure}[h!]
\centering
\begin{overpic}[width=\textwidth]{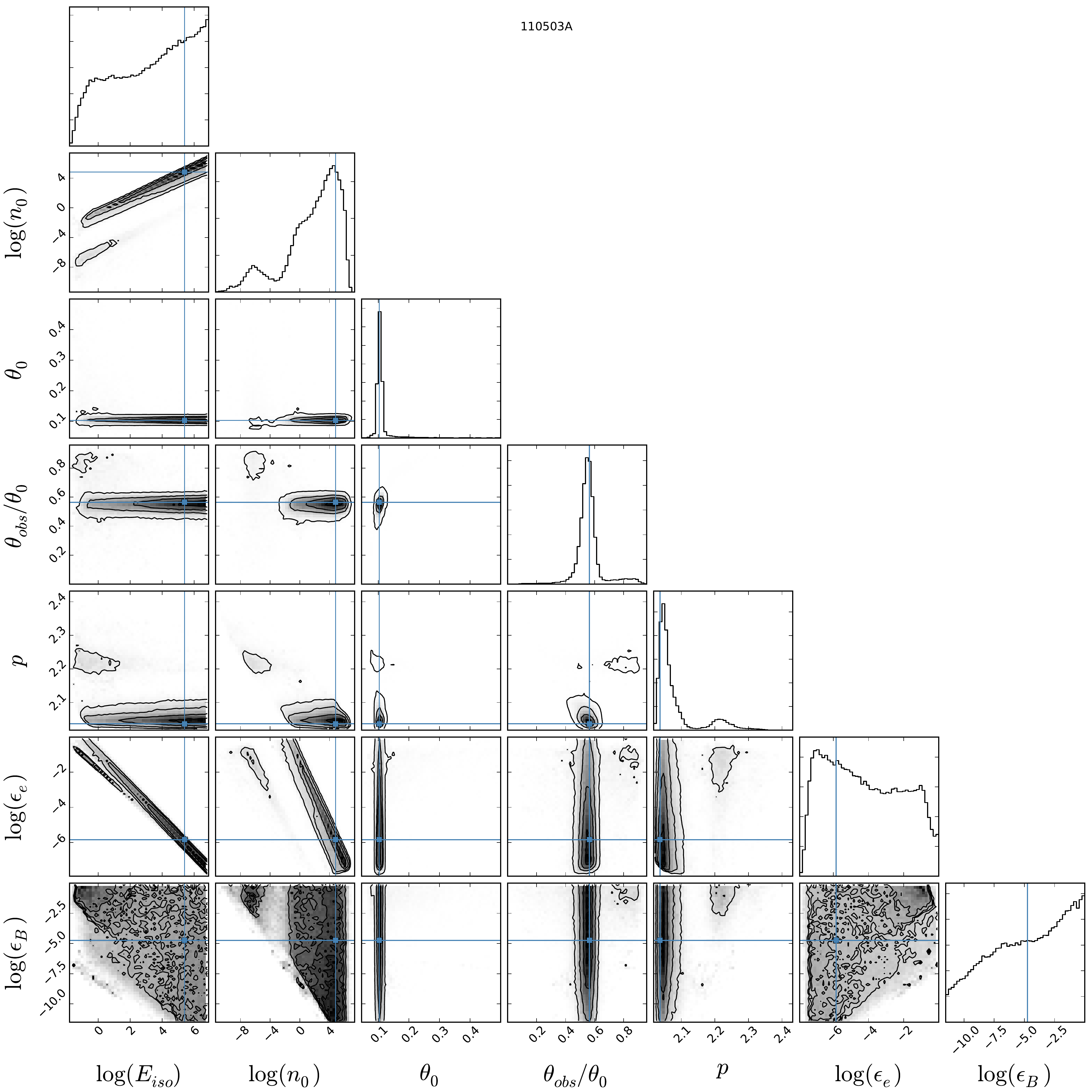}
	\put(46,60){\includegraphics[width=0.55\textwidth]{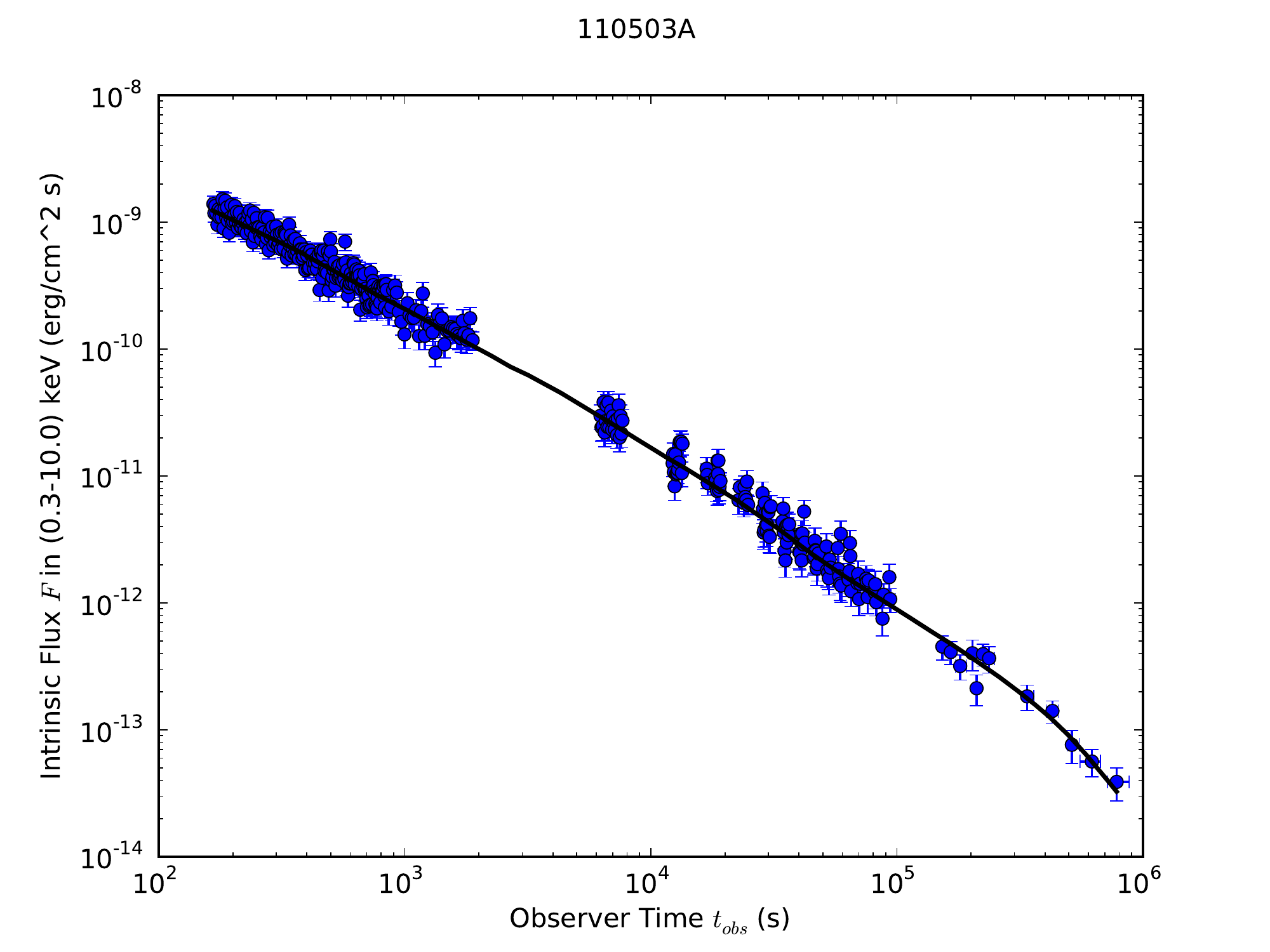}}
\end{overpic}
\caption{Preliminary fit result for 110503A.  The diagonals along the corner plot show the marginalized probabilities for each parameter.  The off-diagonal contour plots show the covariances between all pairs of parameters.  The best-fit values (MAP, maximum posterior probability) are shown in blue.  The best-fit light curve is shown against the data in the upper right.}
\label{fig:fit}
\end{figure}

	For each afterglow in the sample \scalefit{} was run with 1500 random walkers for 2000 iterations, with a burn-in run of 500 iterations.  This produces three million samples of the posterior PDF for each afterglow, covering several auto-correlation times (typically $\sim120$ iterations).  Figure \ref{fig:fit} shows a corner plot of the fit for 110503A, with the marginalized distributions of each parameter along the diagonal and covariance plots in the off-diagonal locations.
	
	The fit for 110503A shows typical behaviour for a well-fit afterglow in our sample.  The values of $\Ei$, $n_0$, and $\epse$ are effectively unconstrained but show an extremely high degree of correlation with each other, as expected from the model.  This is due to a degeneracy in the scaling relations when the afterglow remains in a single spectral regime.  We expect this degeneracy will be removed by performing multi-band fits.  The angles $\thO$ and $\thobs$ as well as the spectral index $p$ are well constrained by the fit, demonstrating the use of this procedure even in single-band fits.  The multi-modality in the $p$ distribution is most likely due to \scalefit{} trying to fit different spectral regimes to the data.  This multi modal behaviour is reflected in the correlations between $p$ and the other parameters, particularly $n_0$.  
	
\begin{figure}[h!]
\centering
\includegraphics[width=0.49\textwidth]{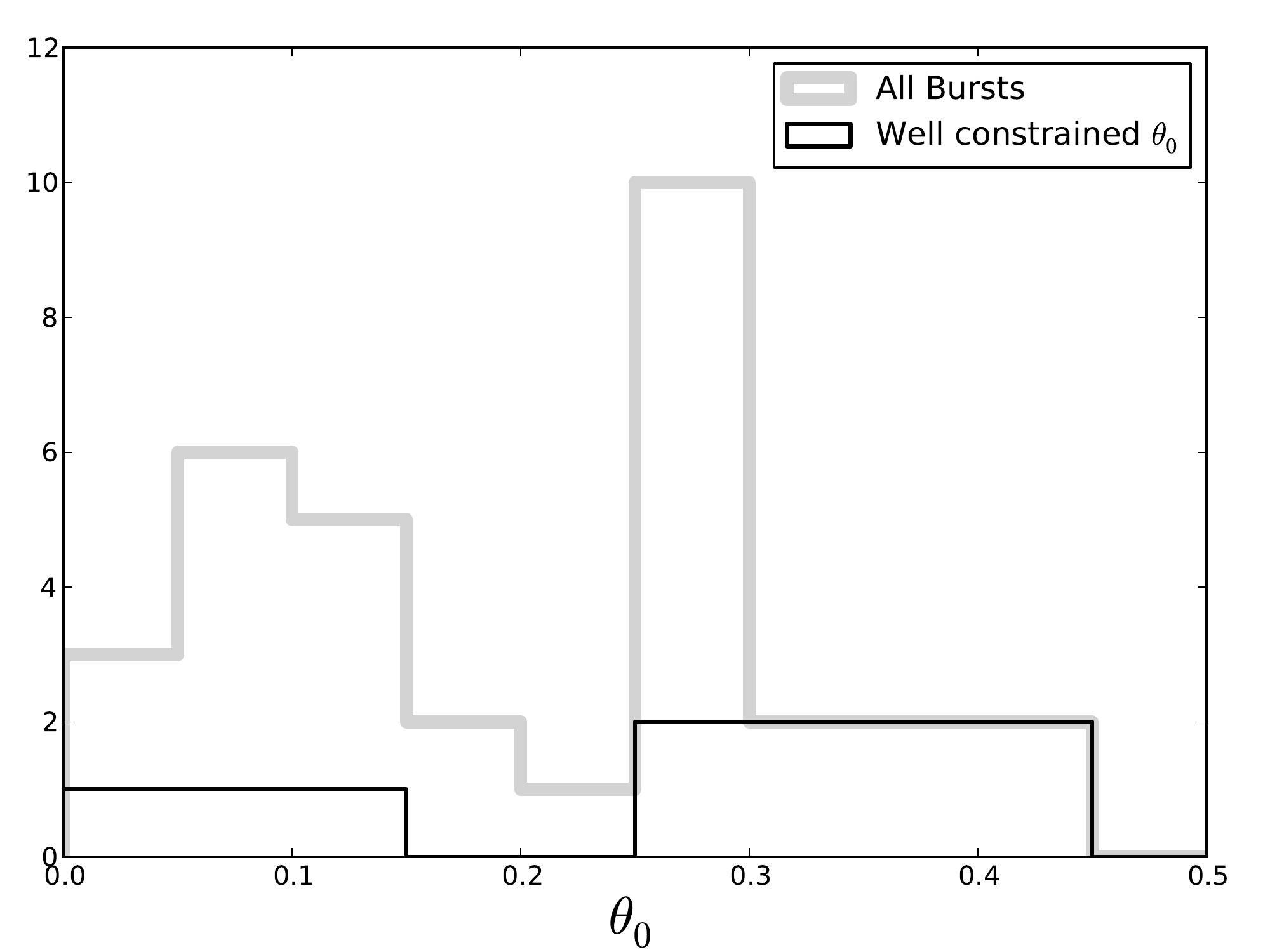}  \includegraphics[width=0.49\textwidth]{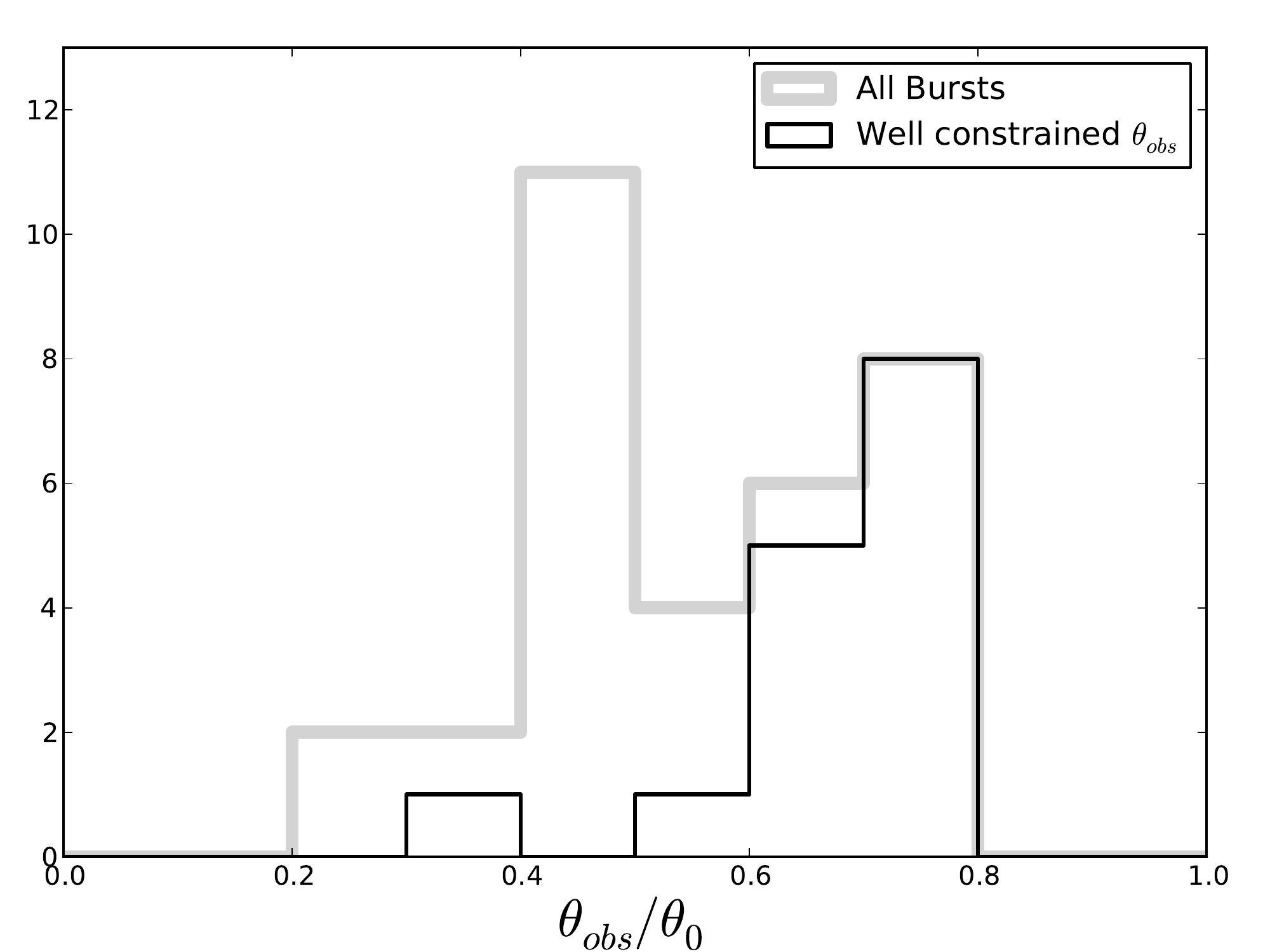} 
\caption{Distribution of median values for $\thO$ and $\thobs$.  Well constrained fits satisfy $\delta \thO / \thO < 0.5$, where $\delta \thO$ is the half-width of the $68\%$ confidence interval.  The same criterion is applied to $\thobs$.  These results are preliminary.}
\label{fig:histthe}
\end{figure}

\begin{figure}[h!]
\centering
 \includegraphics[width=0.49\textwidth]{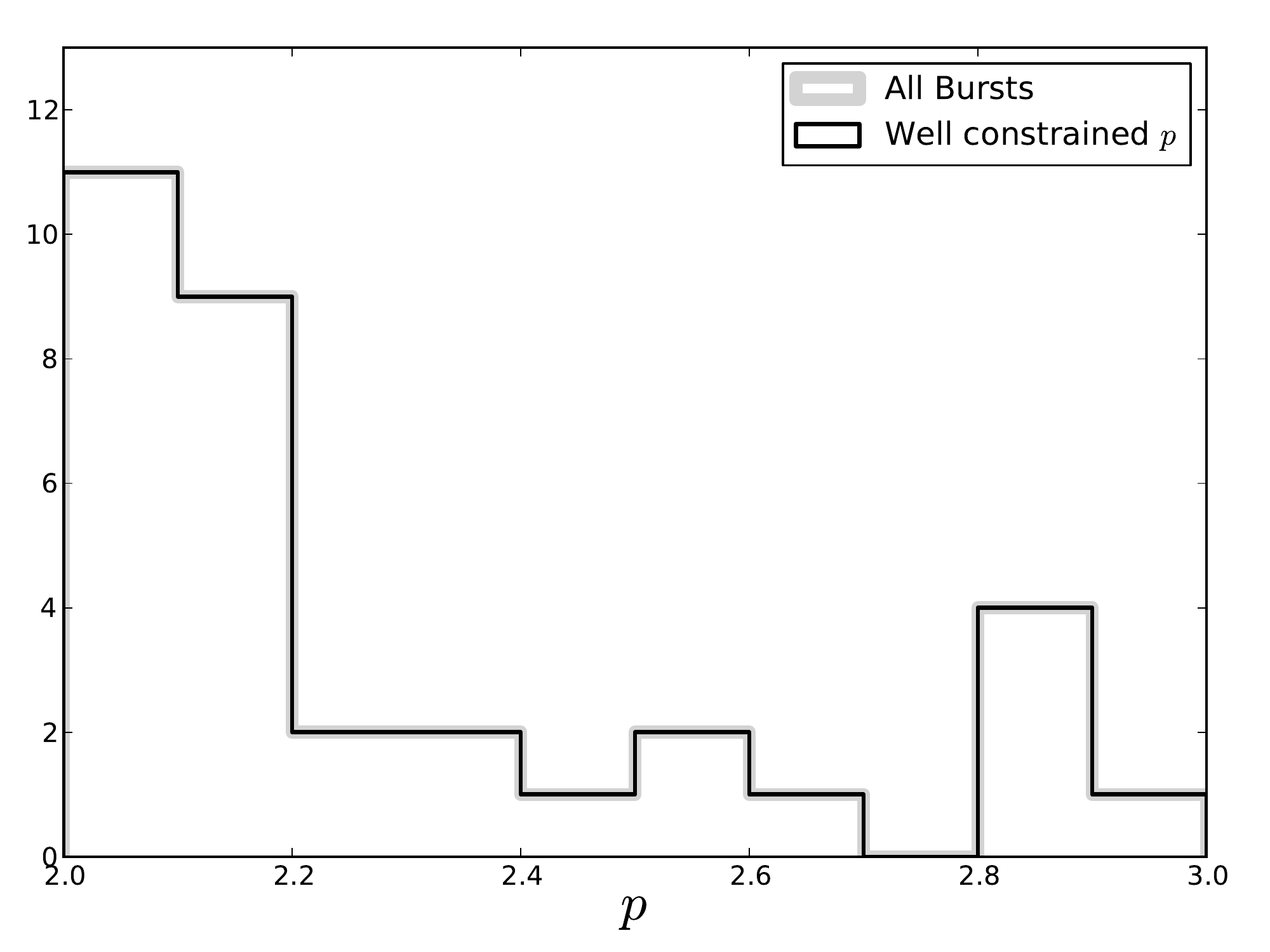}
\caption{Distribution of median values for $p$.  Well constrained fits satisfy $\delta p / p < 0.5$, where $\delta p$ is the half-width of the $68\%$ confidence interval.  All bursts in the sample passed this cut, so the curves are identical.  These results are preliminary.}
\label{fig:histp}
\end{figure}

	Taking our dataset as a representative sample of GRB afterglows we can histogram the central values of a parameter to determine the global distribution for that parameter.  Figures \ref{fig:histthe} and \ref{fig:histp} shows a histogram of median values of $\thO$, $\thobs$, and $p$ over all bursts in our sample.

Several fits resulted in very broad (\ie{} flat) distributions for $\thO$ and $\thobs$.  These distributions tend to have medians near the midpoint of their domain, hence the large number of bursts with $0.25 < \thO < 0.3$ or $0.4 < \thobs / \thO < 0.5$.  In order to cut away these ill-constrained values we make a cut on the fractional uncertainty in each parameter, requiring it to be less than $0.5$.  This cut certainly introduces unknown biases into the resulting distribution, so the distribution over all bursts is plotted as well.  The value for $p$ is determined to good accuracy by all bursts in the sample.  

The well-constrained distribution of $\thO$ is broad, although the all-burst distribution includes a peak at $\thO \sim 0.1$.  The signals from $p$ and $\thobs$ are more clear.  The $p$ distribution favours smaller values, with a large peak (including about 2/3 of bursts in the sample) at $p\sim 2.1$. The observer angle shows a clear preference to be off-axis, peaking at $\thobs \sim 0.7 \thO$.  

Assuming random orientations of GRBs in the sky, the distribution of $\thobs / \thO$ is expected to grow with $\thobs$ until the effects of jet spreading become significant.  Our results corroborate this hypothesis.  This result is striking, as the effect of off-axis observation is not usually included in afterglow fitting.  Off-axis observers see jet-breaks smeared out and completing at later times compared to on-axis observers \cite{vanEerten:2010}.  This affects energy estimates and the determination of opening angles.   

\section{Conclusion}

\scalefit{} is a new data-fitting package for GRB afterglows, allowing high performance MCMC routines to fit afterglow light curves to high resolution, two-dimensional hydrodynamic simulations.  As a first test of the package, we perform single-band fits on all \swift-XRT afterglows with known redshifts in 2011 and 2012.  $\Ei$, $n_0$, $\epse$, and $\epsB$ are difficult to constrain with the single band fit and display a high degree of correlation.  $\thO$, $p$, and the observer angle $\thobs$ are well constrained by the data in several fits, allowing for study of the global distributions of these parameters.  In our preliminary results, we find the afterglow blastwave tend to have $p\sim 2.1$, and be viewed off-axis with $\thobs$ a significant fraction of $\thO$.  \scalefit{} will undergo a future public release.  A more thorough analysis is underway and will appear in a future publication \cite{Ryan:2013}.

\Acknowledgements
This research was supported in part by NASA through grant NNX10AF62G issued
through the Astrophysics Theory Program and by the NSF through grant AST-
1009863 and by the Chandra grant TM3-14005X. Resources supporting this work were
provided by the NASA High-End Computing (HEC) Program through the NASA Ad-
vanced Supercomputing (NAS) Division at Ames Research Center. The software used
in this work was in part developed by the DOE-supported ASCI/Alliance Center for
Astrophysical Thermonuclear Flashes at the University of Chicago.  We thank David Burrows, Binbin Zhang, Judith Racusin, David Hogg, and Daniel Foreman-Mackey for their helpful comments.

\end{document}